\documentclass{article}
\usepackage{amsmath}
\usepackage{epsfig}
\usepackage{amssymb}
\usepackage{fullpage}
\usepackage{subfig}

\def\##1{{\underline #1}}
\def\=#1{\underline{\underline{#1}}}

\def\+#1{\underline{\bf #1}}
\def\*#1{\underline{\underline{\bf #1}}}

\def\eps{\epsilon}
\def\epso{\epsilon_{\scriptscriptstyle 0}}
\def\muo{\mu_{\scriptscriptstyle 0}}
\def\ko{k_{\scriptscriptstyle 0}}
\def\co{c_{\scriptscriptstyle 0}}

\def\.{\mbox{ \tiny{$^\bullet$} }}

\def\ux{\#{u}_x}
\def\uy{\#{u}_y}
\def\uz{\#{u}_z}

\def\le{\left(}
\def\ri{\right)}
\def\les{\left[}
\def\ris{\right]}
\def\lec{\left\{}
\def\ric{\right\}}

\def\l#1{\label{#1}}
\def\r#1{(\ref{#1})}

\def\eps{\epsilon}
\def\epso{\epsilon_0}
\def\muo{\mu_0}
\def\ko{k_0}
\def\co{c_0}

\def\.{\mbox{ \tiny{$^\bullet$} }}

\def\rt{(\#r,t)}

\def\ux{\hat{\#{u}}_x}
\def\uy{\hat{\#{u}}_y}
\def\uz{\hat{\#{u}}_z}
\def\uq{\hat{\#{u}}_q}

\def\uc{\#{u}_c}
\def\Bc{\#{B}_c}
\def\ac{a_c}
\def\epsopc{{\=\eps}^o_c}
\def\epsopj{{\=\eps}^o_j}
\def\epsopcp{{\=\eps}^{o'}_c}
\def\muopc{{\=\mu}^o_c}
\def\muopj{{\=\mu}^o_j}
\def\muopcp{{\=\mu}^{o'}_c}
\def\nuopc{{\=\nu}^o_c}
\def\Copc{\*C^o_c}
\def\Copj{\*C^o_j}
\def\gopc{\*g^o_c}
\def\gop{\*g^o}
\def\gopce{\=g^o_{ce}}
\def\gopcm{\=g^o_{cm}}

\def\Bdc{\#B^{dc}}
\def\Eop{\#E^o}
\def\Bop{\#B^o}
\def\Dop{\#D^o}
\def\Hop{\#H^o}
\def\epsop{{\=\eps}^o}

\def\nuop{{\=\nu}^o}
\def\Cop{\*C^o}

\def\Copuni{\*C^o_{uni}}

\def\eop{\#e^o}

\def\hop{\#h^o}

\def\rn{\#r_n}
\def\rm{\#r_m}
\def\vm{v_m}
\def\vn{v_n}
\def\Vi{V_i}

\def\le{\left(}
\def\ri{\right)}
\def\les{\left[}
\def\ris{\right]}
\def\lec{\left\{}
\def\ric{\right\}}

\def\l#1{\label{#1}}
\def\r#1{(\ref{#1})}

\begin{document}

\begin{center} {\bf {\large Integral Equation for Scattering of Light by a\\
Strong  Magnetostatic Field in Vacuum}}\\
\end{center}

\noindent {\sf AKHLESH LAKHTAKIA} \\
CATMAS~---~Computational \& Theoretical Materials Sciences Group \\
Department of Engineering Science \& Mechanics\\
212 Earth \& Engineering Sciences Building\\
Pennsylvania State University, University Park, PA 16802--6812\\
USA\\

\noindent {\sf TOM G. MACKAY}\\
School of Mathematics\\
James Clerk Maxwell Building\\
University of Edinburgh\\
Edinburgh EH9 3JZ\\
UK


\begin{abstract} When a strong magnetostatic field is present,
vacuum effectively appears as a linear, uniaxial, dielectric--magnetic
medium for small--magnitude optical fields. The availability of the
frequency--domain dyadic Green function when the magnetostatic field is
spatially uniform facilitates the formulation  of
an integral equation for the scattering of an optical field by a
spatially varying magnetostatic field in vacuum. This integral
equation can be numerically treated by using the method of moments
as well as the coupled dipole method. Furthermore, the principle
underlying the strong--property--fluctuation theory allows the
homogenization of a spatially varying magnetostatic field in the
context of light scattering.

\end{abstract}

\section{Introduction}
In classical electrodynamics,
light propagating in vacuum (i.e., matter--free space) is not considered to be
affected by the presence of a magnetostatic field. This is because
classical vacuum  is a linear  medium wherein the principle of superposition holds. But, in quantum electrodynamics (QED), vacuum is a nonlinear medium (Jackson, 1998). It can, however,
be linearized for a rapidly time--varying electromagnetic field with a small
amplitude in the presence of a slowly varying (or static) magnetic field (Adler, 2007). The price of linearization is that the QED vacuum appears as an anisotropic dielectric--magnetic medium for optical fields (Adler, 1971).

Our modest aim in this paper is to derive an integral equation for
the scattering of a high--frequency electromagnetic field (typified
by light) by a strong magnetostatic field in vacuum. For this
purpose, we exploit the analytical machinery developed during the
last two decades for the frequency--domain analysis of
electromagnetic fields in complex mediums (Singh \& Lakhtakia, 2000;
Weiglhofer \& Lakhtakia, 2003). As the derived integral equation
shall have to be solved numerically in general, we also outline the
solution strategies provided by the method of moments (Miller,
Medgyesi--Mitschang, \& Newman, 1991; Wang, 1991) and the coupled
dipole method (Purcell \& Pennypacker, 1973; Lakhtakia, 1990).
Finally, we adopt the principle underlying the
strong--property--fluctuation theory to homogenize the magnetostatic
field in the context of light scattering.

A note about notation:  3--vectors (6--vectors) are in normal (bold)
face and underlined, whereas 3$\times$3 dyadics (6$\times$6 dyadics)
are in normal (bold) face and double underlined.  The position
vector is denoted by $\#r=x\ux+y\uy+z\uz$ in a Cartesian coordinate
system with unit vectors $\ux$, $\uy$, and $\uz$, whereas time is
denoted by $t$. The real part of a complex--valued quantity $\zeta$ is
written as ${\Re} \lec \zeta \ric$.

\section{Constitutive Equations for Optical Fields}

Suppose that all space is matter--free and that a magnetostatic field $\Bdc(\#r)$
is present everywhere. Optical fields (superscripted ``o" in this
paper) satisfy the source--free Maxwell equations (Adler, 1971)
\begin{equation}
\left.
\begin{array}{l}
\nabla\cdot\Bop\rt=0\\[5pt]
\nabla\cdot\Dop\rt=0\\[5pt]
\nabla\times\Eop\rt=-\frac{\partial}{\partial t} \,\Bop\rt\\[5pt]
\nabla\times\Hop\rt=\frac{\partial}{\partial t} \,\Dop\rt
\end{array}
\right\}\,,
\label{MaxEqopt}
\end{equation}
provided $\vert\Bop\vert\ll\vert\Bdc(\#r)\vert\,\forall\#r$. As per the linearized version of
QED, the optical fields appearing in the
foregoing equations obey the constitutive equations (Adler,1971)
\begin{equation}
\left.
\begin{array}{l}
\Dop\rt=\epso\,\epsop(\#r)\cdot\Eop\rt\\[5pt]
\Hop\rt=\muo^{-1}\,\nuop(\#r)\cdot\Bop\rt
\end{array}
\right\}\,,
\end{equation}
where $\epso=8.8542\times10^{-12}$~F~m$^{-1}$ and $\muo=4\pi\times 10^{-7}$~H~m$^{-1}$ are constants characterizing the classical vacuum, whereas the relative
permittivity dyadic
\begin{equation}
\epsop(\#r) = \les1-8\epso\co^2\xi\,\Bdc(\#r)\cdot\Bdc(\#r)\ris\=I +28\epso\co^2\xi\,\Bdc(\#r)\Bdc(\#r)
\end{equation}
and the  relative impermeablity dyadic\footnote{Impermeability is the reciprocal
of the permeability.}
\begin{equation}
\nuop(\#r) = \les1-8\epso\co^2\xi\,\Bdc(\#r)\cdot\Bdc(\#r)\ris\=I -16\epso\co^2\xi\,\Bdc(\#r)\Bdc(\#r)
\end{equation}
emerge from QED to dictate the influence of the
magnetostatic field of high magnitude  on the optical field. Here, $\co=1/\sqrt{\epso\muo}$ is the speed of light in
classical vacuum, whereas
\begin{equation}
\xi=\frac{(e_{e\ell}/m_{e\ell})^4\hbar}{45 (4\pi \epso)^2 \co^7}
= 8.3229\times 10^{-32}\,\, {\mbox {kg}}^{-1}\,\,{\mbox{m}}\,\,{\mbox{s}}^2
\end{equation}
contains the electronic charge $e_{e\ell}=1.6022\times 10^{-19}$~C, the electronic mass
$m_{e\ell}=9.1096\times 10^{-31}$~kg, and the reduced Planck constant
$\hbar=1.0546\times10^{-34}$~J~s. Clearly, the QED vacuum appears
to the optical field as a spatiotemporally local, spatially nonhomogeneous, temporally unvarying,
uniaxial dielectric--magnetic medium. Set $\hbar=0$ to convert from the QED vacuum to the classical vacuum, and the influence of the magnetostatic field on the optical field
vanishes.

\section{Dyadic Green Function for Uniform Magnetostatic Field}
Suppose that $\Bdc(\#r) = \Bc$ is spatially uniform everywhere, so that the QED vacuum
then is spatially homogeneous. Defining
\begin{equation}
\left.
\begin{array}{l}
\ac=\epso\co^2\xi\,\vert\Bc\vert^2\\[5pt]
\uc= \Bc/\vert\Bc\vert
\end{array}
\right\}\,,
\end{equation}
we can write the relative permittivity dyadic
\begin{equation}
\label{epsopcdef}
\epsopc=(1-8\ac)(\=I-\uc\uc) + (1+20\ac)\,\uc\uc
\end{equation}
and the relative permeability dyadic
\begin{equation}
\label{muopcdef}
\muopc=\left(\nuopc\right)^{-1}
=\frac{1}{1-8\ac}\,(\=I-\uc\uc) +\frac{1}{1-24\ac}\,\uc\uc\,
\end{equation}
for the QED vacuum, with $\=I$ as the 3$\times$3 identity dyadic.
 The propagation of optical plane waves,
characterized by
\begin{equation}
\Eop\rt = {\Re}\lec \tilde{\Eop}\, \exp\left[i(\#k\cdot\#r-\omega t)\right]\ric\,,
\end{equation}
etc., with amplitude vector $\tilde{\Eop}$, wave vector $\#k$, and angular frequency $\omega$ was analyzed
by Adler (1971) using standard mathematical techniques. Both the relative
permittivity and the relative permeability dyadics are uniaxial and share the same
distinguished axis. Therefore, the medium is optically birefringent except when
$\#k$ is parallel to $\uc$; furthermore, both plane waves propagating in
a fixed direction are to be classified as \emph{extraordinary}
in optical parlance (Lakhtakia, Varadan, \& Varadan, 1991).

For a more general consideration of frequency--domain
optical fields, it is best to write
\begin{equation}
\label{eopdef}
\Eop\rt = {\Re}\lec \eop(\#r) \exp(-i\omega t)\ric\,,
\end{equation}
etc., where $\eop(\#r)$ is a phasor.
The Maxwell curl equations may then be compactly recast with the help of
6--vectors and 6$\times$6 dyadics as
\begin{equation}
\label{6diffeq}
\les\*L(\nabla) + i\omega \Copc\ris\cdot \+f^o(\#r) = \+s^o(\#r)\,,
\end{equation}
where
\begin{equation}
\*L(\nabla) =
\les \begin{array}{cc}
\=0 & \nabla\times\=I\\[5pt]
-\nabla\times\=I & \=0
\end{array}\ris\,,
\end{equation}
\begin{equation}
\label{Copcdef}
\Copc =
\les \begin{array}{cc}
\epso\,\epsopc & \=0\\[5pt]
\=0 & \muo\,\muopc
\end{array}\ris\,,
\end{equation}
\begin{equation}
\+f^o(\#r)=\les\begin{array}{c}
\eop(\#r)\\[5pt]\hop(\#r)\end{array}\ris\,,
\end{equation}
$\+s^o(\#r)$ represents the (high--frequency) electric and magnetic source
current  densities
that engender $\+f^o(\#r)$, and $\=0$ is the 3$\times$3 null dyadic.

Equation \r{6diffeq} is converted into the integral equation
\begin{equation}
\label{6diffeqsol}
\+f^o(\#r) = \+f^o_h(\#r) + \int d^3\#r'\, \gopc(\#r,\#r')\cdot \+s^o(\#r')\,,
\end{equation}
where $\+f^o_h(\#r)$ is the solution of \r{6diffeq} when the source term on its
right side is null--valued {\it everywhere}. The 6$\times$6 dyadic Green function
$\gopc(\#r-\#r')$ is the solution of
\begin{equation} \l{goc_def}
\les\*L(\nabla) + i\omega \Copc\ris\cdot\gopc(\#r,\#r') = \*I\,\delta(\#r-\#r')\,,
\end{equation}
where $\*I$ is the 6$\times$6 identity dyadic and $\delta(\#r-\#r')$ is
Dirac delta function.

The 6$\times$6 dyadic Green function
$\gopc(\#r,\#r')$ is available from the literature on electromagnetic fields
in uniaxial mediums as (Weiglhofer, 1990)
\begin{equation}
\gopc(\#r,\#r')=\les
\begin{array}{cc}
\=I & -(i\omega \epso\,\epsopc)^{-1}\cdot(\nabla\times\=I)\\[5pt]
(i\omega\muo\muopc)^{-1}\cdot(\nabla\times\=I) & \=I
\end{array}\ris\,
\les\begin{array}{cc}
\gopce(\#r,\#r') & \=0 \\[5pt]
\=0 & \gopcm(\#r,\#r')\end{array}
\ris\,,
\end{equation}
where the 3$\times$3 dyadic Green functions
\begin{equation}
\gopce(\#r,\#r')=-(1-8\ac)^{-1}\lec   (i\omega\epso)^{-1}
\les \nabla\nabla +\ko^2(1+20\ac)(\epsopc)^{-1}\ris\,\gamma^o_e(\#r,\#r')
+i\omega\muo\,\=\varphi(\#r,\#r')\ric\,
\end{equation}
and
\begin{equation}
\gopcm(\#r,\#r')=-(1-8\ac)\lec   (i\omega\muo)^{-1}
\les \nabla\nabla +\ko^2(1-24\ac)^{-1}(\muopc)^{-1}\ris\,\gamma^o_m(\#r,\#r')
-i\omega\epso\,\=\varphi(\#r,\#r')\ric\,
\end{equation}
employ the usual wavenumber $\ko=\omega(\epso\muo)^{1/2}$
for classical vacuum. The 3$\times$3 dyadic function
\begin{eqnarray}
\nonumber
&&\=\varphi(\#r,\#r')=
\les \left(\frac{1+20\ac}{1-8\ac}\right)\gamma^o_e(\#r,\#r')
- \left(\frac{1-8\ac}{1-24\ac}\right)\gamma^o_m(\#r,\#r')\ris
\frac{(\#R\times\uc)(\#R\times\uc)}{\vert\#R\times\uc\vert^2}\
\\[5pt]
&&\qquad+\les \=I -\uc\uc-2\frac{(\#R\times\uc)(\#R\times\uc)}{\vert\#R\times\uc\vert^2}\ris
\frac{R^o_e\,\gamma^o_e(\#r,\#r')-R^o_m\,\gamma^o_m(\#r,\#r')}
{i\ko\vert\#R\times\uc\vert^2}\,
\end{eqnarray}
and the scalar functions
\begin{equation}
\gamma^o_e(\#r,\#r')=\frac{\exp(i\ko R^o_e)}{4\pi R^o_e}\,
\end{equation}
and
\begin{equation}
\gamma^o_m(\#r,\#r')=\frac{\exp(i\ko R^o_m)}{4\pi R^o_m}\,
\end{equation}
contain
\begin{eqnarray}
&&\#R=\#r-\#r'\,,\\
&&R^o_e=+\les
 \left(\frac{1+20\ac}{1-8\ac}\right)\vert\#R\times\uc\vert^2+(\#R\cdot\uc)^2\ris^{1/2}
 \,,\\[5pt]
 &&R^o_m=+\les
 \left(\frac{1-8\ac}{1-24\ac}\right)\vert\#R\times\uc\vert^2+(\#R\cdot\uc)^2\ris^{1/2}\,.
\end{eqnarray}

\section{Formulation of Integral Equation}
Returning to the more general case of a nonuniform magnetostatic field $\Bdc(\#r)$,
and making use of the representation
\r{eopdef}, we see
that
the two curl equations in \r{MaxEqopt} may be written together as
\begin{equation}
\label{6diffeq1}
\les\*L(\nabla) + i\omega \Cop(\#r)\ris\cdot \+f^o(\#r) = \+0\,,
\end{equation}
where $\+0$ is the null 6--vector and
\begin{equation}
\Cop (\#r) =
\les \begin{array}{cc}
\epso\,\epsop(\#r) & \=0\\[5pt]
\=0 & \muo\,\les\nuop(\#r)\ris^{-1}
\end{array}\ris\,.
\end{equation}
As a simple Green--function formalism is unlikely to be found for this more general
case, we \emph{define} a suitable uniform magnetostatic field $\Bc$ and proceed as follows.

\subsection{Derivation}
Equation \r{6diffeq1} is recast as
\begin{equation}
\label{6diffeq2}
\les\*L(\nabla) + i\omega \Copc\ris\cdot \+f^o(\#r) =\+s^o_{eq}(\#r)\,,
\end{equation}
where
\begin{equation}
\+s^o_{eq}(\#r)=
i\omega \les\Copc- \Cop(\#r)\ris\cdot \+f^o(\#r)
\end{equation}
is an \emph{equivalent} high--frequency source
current density 6--vector that contains the spatial variations
of $\Bdc(\#r)$ in relation to $\Bc$. Clearly, $\Bc$ should be chosen carefully. If the variations of $\Bdc(\#r)$ are confined to some bounded region only, $\Bc$ should
be chosen as the value of $\Bdc(\#r)$ outside that region. If $\Bdc(\#r)$ varies
over all space, $\Bc$ may emerge from some homogenization or spatial--averaging procedure. Once that choice has been
made, the solution of \r{6diffeq2}~---~and therefore of \r{6diffeq1}~---~may be written as
\begin{equation}
\label{6diffeqsol2}
\+f^o(\#r) = \+f^o_h(\#r) + i\omega\int d^3\#r'\, \gopc(\#r,\#r')\cdot \les\Copc- \Cop(\#r')\ris\cdot \+f^o(\#r')\,,
\end{equation}
where $\+f^o_h(\#r)$ is the homogeneous part of the solution of
\r{6diffeq2}.

The integrand on the right side of \r{6diffeqsol2} is singular at $\#r=\#r'$ and therefore
requires additional treatment.
An ellipsoidal region $V_\varepsilon$ containing the location
$\#r$ midway between its two focuses is identified in the domain of integration. The surface of $V_\varepsilon$ is the set of
points
\begin{equation}
\#r_\varepsilon(\theta_q,\phi_q)=\#r + \varepsilon \,\=U\cdot\uq\,,\quad
\theta_q\in\les0,\pi\ris\,,\quad\phi_q\in\left[0,2\pi\right]\,,
\end{equation}
where $\varepsilon$ is a positive scalar, the unit vector
\begin{equation}
\uq = (\ux\,\cos\phi_q +\uy\,\sin\phi_q)\sin\theta_q+\uz\,\cos\theta_q\,,
\end{equation}
and the dyadic
\begin{equation}
\=U=a_x\,\ux\ux +a_y\,\uy\uy +a_z\,\uz\uz\,
\end{equation}
has positive eigenvalues $a_x$, $a_y$, and $a_z$, all less than or equal
to unity. The integral is divided into two
parts: the first is to be evaluated over  $V_\varepsilon$ only, the second over all space except
$V_\varepsilon$. Both parts are evaluated in the limit $\varepsilon\to0$ in order to get the desired
integral equation
\begin{equation}
\label{6diffeqsol3}
\+f^o(\#r) = \+f^o_h(\#r) + i\omega\*D\cdot \les\Copc- \Cop(\#r)\ris\cdot \+f^o(\#r)
+ i\omega\,{\sf PV}\int d^3\#r'\, \gopc(\#r,\#r')\cdot \les\Copc- \Cop(\#r')\ris\cdot \+f^o(\#r')
\,,
\end{equation}
wherein the symbol $\sf PV$ identifies that the integral following it has to be
evaluated
in the `principal--value sense', as described.

\subsection{Depolarization Dyadic}
The 6$\times$6 depolarization dyadic $\*D$ in \r{6diffeqsol3} is a two--dimensional
surface integral
(Michel \& Weiglhofer 1997; Weiglhofer, Lakhtakia, \& Michel 1997):
\begin{eqnarray} \l{DD_def}
\*D &=& -\,\frac{i\omega}{4\pi  \ko^2} \, \Bigg\{
\int_0^{2\pi}d\phi_q \int_0^\pi d\theta_q\, \sin\theta_q \nonumber
\\ && \times \les\begin{array}{cc}
\muo\, \=U^{-1}\cdot \frac{(\uq\cdot\=U^{-1})(\uq\cdot\=U^{-1})}{(\uq\cdot\=U^{-1})\cdot\epsopc\cdot(\uq\cdot\=U^{-1})}\cdot \=U^{-1}&\=0\\[8pt]
\=0 & \epso\, \=U^{-1}\cdot
\frac{(\uq\cdot\=U^{-1})(\uq\cdot\=U^{-1})}{(\uq\cdot\=U^{-1})\cdot\muopc\cdot(\uq\cdot\=U^{-1})}\cdot
\=U^{-1}
\end{array}\ris \Bigg\} .
\end{eqnarray}
In general, it has to be evaluated numerically, which can accomplished
quite easily by using the Gauss--Legendre quadrature scheme, for
instance (Mackay \& Weiglhofer, 2000).

In two special situations, $\*D$ can be obtained analytically. Suppose, first,
that the ellipsoidal region $V_\varepsilon$ is spherical. Then, $\=U=\=I$ so that
\begin{equation}
\label{Dfirst}
\*D=
-\,\frac{i\omega}{4\pi  \ko^2}\int_0^{2\pi}\,d\phi_q\,\int_0^\pi\,d\theta_q\,
\sin\theta_q
\les\begin{array}{cc}
\muo\,\frac{\uq\uq}{\uq\cdot\epsopc\cdot\uq}&\=0\\[5pt]
\=0 &\epso\, \frac{\uq\uq}{\uq\cdot\muopc\cdot\uq}
\end{array}\ris\,.
\end{equation}
For $\epsopc$ defined in \r{epsopcdef} and $\muopc$ defined in \r{muopcdef},
the two integrals on the right side of \r{Dfirst} can be evaluated in closed form
to yield (Mackay \& Lakhtakia 2005)
\begin{eqnarray}
\nonumber
&&
\*D= -\,\frac{i\omega}{4\pi  \ko^2} \,
\les\begin{array}{cc}
 \frac{\muo}{ 1 - 8 \ac }   &\=0\\[5pt]
\=0 & \epso ( 1 - 8 \ac )
\end{array}\ris
\\[5pt]
&&\qquad\times
\les\begin{array}{cc}
 \Gamma_t (\rho^\eps) \le
\=I-\uc\uc \ri + \Gamma_c (\rho^\eps)\,\uc\uc  &\=0\\[5pt]
\=0 &   \Gamma_t (\rho^\mu) \le
\=I-\uc\uc \ri + \Gamma_c (\rho^\mu) \,\uc\uc
\end{array}\ris
\end{eqnarray}
where
\begin{eqnarray}
\Gamma_t ( \rho )&=& \left\{
\begin{array}{lcr}
\displaystyle{ \frac{1}{2} \le  \frac{1}{1-\rho }-
  \frac{ \rho \sinh^{-1} \sqrt{\frac{1
-\rho}{\rho} }}{\le 1 - \rho  \ri^{\frac{3}{2}}} \ri } && \mbox{for}
\;\; 0 < \rho < 1
\\ & & \\
\displaystyle{\frac{1}{2} \le \frac{\rho \sec^{-1} \sqrt{\rho} }
{\le \rho - 1 \ri^{\frac{3}{2}}} - \frac{1}{\rho - 1}  \ri }& &
\mbox{for} \;\; \rho > 1
\end{array}
\right., \\
\Gamma_c (\rho )&=& \left\{
\begin{array}{lcr}
\displaystyle{
  \frac{\sinh^{-1} \sqrt{\frac{1
-\rho}{\rho} }}{\le 1 - \rho  \ri^{\frac{3}{2}}} - \frac{1}{1-\rho
}} && \hspace{14mm} \mbox{for} \;\; 0 < \rho < 1
\\ & & \\
\displaystyle{ \frac{1}{\rho - 1} - \frac{\sec^{-1} \sqrt{\rho} }
{\le \rho - 1 \ri^{\frac{3}{2}}}}& & \mbox{for} \;\; \rho > 1
\end{array}
\right., \\
\end{eqnarray}
and
\begin{equation}
\left.
\begin{array}{l}
\rho^\eps = \frac{1 + 20 \ac}{1- 8 \ac}\vspace{12pt}\\
\rho^\mu = \frac{1 - 8 \ac}{1- 24 \ac}
\end{array}
\right\}.
\end{equation}

The second special situation comes when one of the semi--axes of
the ellipsoidal region $V_\varepsilon$ coincides with $\uc$. Without loss of
generality, let us choose  $\uc = \hat{\#u}_z$ and then introduce the
3$\times$3 dyadics
\begin{equation}
\left. \begin{array}{l} \epsopcp = \=U^{-1} \cdot \epsopc \cdot
\=U^{-1} = \eps^{o'}_x\,\ux\ux +\eps^{o'}_y\,\uy\uy +\eps^{o'}_z\,\uz\uz \vspace{8pt} \\
\muopcp = \=U^{-1} \cdot \muopc \cdot \=U^{-1} = \mu^{o'}_x\,\ux\ux
+\mu^{o'}_y\,\uy\uy +\mu^{o'}_z\,\uz\uz
\end{array}
\right\},
\end{equation}
wherein
\begin{equation}
\left.
\begin{array}{l}
\eps^{o'}_x = \frac{1 - 8 \ac}{a^2_x} \vspace{6pt} \\
\eps^{o'}_y = \frac{1 - 8 \ac}{a^2_y} \vspace{6pt} \\
\eps^{o'}_z = \frac{1 + 20 \ac}{a^2_z}
\end{array}
\right\}\,,\qquad
\left.
\begin{array}{l}
\mu^{o'}_x = \frac{1}{\le 1 - 8 \ac \ri a^2_x} \vspace{6pt} \\
\mu^{o'}_y = \frac{1}{\le 1 - 8 \ac \ri a^2_y} \vspace{6pt} \\
\mu^{o'}_z = \frac{1}{\le 1 - 24 \ac \ri a^2_z}
\end{array}
\right\}.
\end{equation}
It then transpires that the depolarization dyadic may be represented
as (Weiglhofer, 1998)
\begin{equation}
\*D= -\,\frac{i\omega}{4\pi  \ko^2} \,\les\begin{array}{cc}
\muo \, \=U^{-1} \cdot \=d^\eps \cdot \=U^{-1}  &\=0\\[5pt]
\=0 & \epso \,  \=U^{-1} \cdot \=d^\mu \cdot \=U^{-1}
\end{array}\ris\,,
\end{equation}
where
\begin{equation}
 \=d^\eta = d^\eta_x\,\ux\ux + d^\eta_y\,\uy\uy +d^\eta_z\,\uz\uz\,, \qquad ( \eta = \eps, \mu ),
\end{equation}
with
\begin{equation}
\left.
\begin{array}{l}
d^\eta_x = \displaystyle{\frac{\le \eta^{o'}_{y}\ri^{1/2}  \,  \les
F(\lambda_1, \lambda_2) - E(\lambda_1,\lambda_2) \ris}{\le
\eta^{o'}_{y} - \eta^{o'}_{x}    \ri \, \le \eta^{o'}_{z}   -
\eta^{o'}_{x} \ri^{1/2}}} \,
\\[-2mm]
\\
d^\eta_y = \displaystyle{\frac{1}{\eta^{o'}_{y}    - \eta^{o'}_{x}
}}\, \Bigg\{ \, \displaystyle{\frac{\eta^{o'}_{x}    - \eta^{o'}_{y}
 }{\eta^{o'}_{z}    - \eta^{o'}_{y}  }} -
\displaystyle{\le \, \frac{\eta^{o'}_{z}    - \eta^{o'}_{x}
}{\eta^{o'}_{y}
 } \, \ri^{1/2}}  \\ \hspace{11mm} \times \Bigg[ \,
\displaystyle{\frac{\eta^{o'}_{x}   }{\eta^{o'}_{z}
  - \eta^{o'}_{x}  }} \, F(\lambda_1,\lambda_2) -
  \displaystyle{\frac{\eta^{o'}_{y}   }{\eta^{o'}_{z}    - \eta^{o'}_{y}   }}\, E(\lambda_1,\lambda_2) \,
  \Bigg]
\Bigg\}
\\[-2mm]
\\
d^\eta_z = \displaystyle{\frac{1}{\eta^{o'}_{z}    - \eta^{o'}_{y}
}}\, \lec \, 1 - \displaystyle{\le \frac{\eta^{o'}_{y}
}{\eta^{o'}_{z} - \eta^{o'}_{x}
  } \, \ri^{1/2}} \, E(\lambda_1,\lambda_2)\, \ric
\end{array}
\right\}\,,  \quad ( \eta = \eps, \mu )\,,
\end{equation}
which involve $F(\lambda_1,\lambda_2)$ and $E(\lambda_1,\lambda_2)$
as elliptic integrals of the first and second kinds (Gradshteyn \& Ryzhik, 1980),
respectively, with arguments
\begin{equation}
\left.
\begin{array}{l}
 \lambda_1 = \tan^{-1} \le \, \displaystyle{\frac{\eta^{o'}_{z}   -
\eta^{o'}_{x}   }{\eta^{o'}_{x}  } }\, \ri^{1/2}\, \\ \\
\lambda_2= \les \displaystyle{\frac{\eta^{o'}_{z}   \le \,
\eta^{o'}_{y}
  - \eta^{o'}_{x}   \, \ri} {\eta^{o'}_{y}    \le \,
\eta^{o'}_{z}   - \eta^{o'}_{x}   \, \ri}} \, \ris^{1/2}\,
\end{array}
\right\}\,,
 \qquad \quad ( \eta = \eps, \mu )\,.
\end{equation}

\subsection{Numerical--Solution Techniques}
As may be easily guessed, \r{6diffeqsol3} shall have to be solved numerically. When
the spatial variations of $\Bdc(\#r)$ are confined to some bounded region
$\Vi$ whereas the magnetostatic field is uniform with a value $\Bc$ everywhere outside $\Vi$, the
method of moments (Miller, Medgyesi--Mitschang, \& Newman, 1991; Wang, 1991) and the coupled dipole method
(Purcell \& Pennypacker, 1973; Lakhtakia,1990) offer relatively easy algorithms
to implement. Both methods are related to each other (Lakhtakia, 1992), and their adaptations for the present purposes are described as follows.

\subsubsection{Method of Moments}
Implementation of the method of moments requires that $\Vi$ be replaced
by a lattice of points $\rn$, $n\in[1,N]$. Attached to every $r_n$ is an
electrically small region described by a shape dyadic $\=U_n$
 and of volume $v_n$; the sum $(\sum_{n=1}^N\,\vn)$ equals
the volume of $V_i$. Furthermore, a depolarization dyadic $\*D_n$ is associated
with every $\rn$.

Equation \r{6diffeqsol3} is specialized to $\#r=\rn$, $n\in[1,N]$, to obtain the set of
algebraic equations
\begin{eqnarray}
\nonumber
&&
\+f^o(\rn) \approxeq \+f^o_h(\rn)+i\omega\*D_n\cdot\les \Copc-\Cop(\rn)\ris\cdot
\+f^o(\rn)\\
&&\qquad\qquad+\sum_{m=1,m\ne n}^{N}\,
\vm\, \gopc(\rn,\rm)\cdot\les \Copc-\Cop(\rm)\ris\cdot
\+f^o(\rm)\,,\quad n\in[1,N]\,,
\label{mom-1}
\end{eqnarray}
where $\+f^o_h(\#r)$ represents the incident optical field (i.e., the optical fields
when $\Bdc(\#r)=\Bc\,\forall \#r\in \Vi$). This set of equations is rewritten
as
\begin{equation}
\+f^o_h(\rn)=\sum_{m=1}^{N}\, \*Q^{MOM}_{nm}\cdot\+f^o(\rm)\,,\qquad
n\in[1,N]\,,
\label{mom-2}
\end{equation}
where
\begin{equation}
\label{Qmom}
\*Q^{MOM}_{nm} =\lec\*I-i\omega\*D_n\cdot\les \Copc-\Cop(\rn)\ris\ric\delta_{nm}-\vm\gopc(\rn,\rm)\cdot\les\Copc-\Cop(\rm)\ris(1-\delta_{nm})\,,
\end{equation}
with $\delta_{nm}$ as the Kronecker delta function.

Equations \r{mom-2} are solved \emph{for all}
$\+f^o(\rn)$, $n\in[1,N]$ by a
variety of numerical techniques (Carnahan, Luther, \& Wilkes, 1969), with the conjugate gradient method
being the preferred technique when $N$ is large (Strang, 1986; Sarkar, 1991).
Once all $\+f^o(\rn)$ are known, the total field at any $\#r\notin \Vi$ may be determined
from \r{6diffeqsol3} as
\begin{equation}
\+f^o(\#r) \approxeq \+f^o_h(\#r)+\sum_{m=1}^{N}\,
\vm\, \gopc(\#r,\rm)\cdot\les \Copc-\Cop(\rm)\ris\cdot
\+f^o(\rm)\,,\quad\#r\notin \Vi\,.
\label{mom-3}
\end{equation}
The difference $\+f^o(\#r) - \+f^o_h(\#r)$ yields the scattered optical field at
$\#r\notin \Vi$.

\subsubsection{Coupled Dipole Method}
In the coupled dipole method, the first and the third terms on the right side of
\r{mom-1} are used to define an \emph{exciting} optical field via
\begin{equation}
\+f^o_{exc}(\rn) \approxeq \+f^o_h(\rn)+\sum_{m=1,m\ne n}^{N}\,
\vm\, \gopc(\rn,\rm)\cdot\les \Copc-\Cop(\rm)\ris\cdot
\+f^o(\rm)\,,\quad n\in[1,N]\,.
\label{cdm-1}
\end{equation}
Thereafter, the set of equations \r{mom-1} is rewritten as
\begin{equation}
\+f^o(\rn) \approxeq \+f^o_{exc}(\rn)+i\omega\*D_n\cdot\les \Copc-\Cop(\rn)\ris\cdot
\+f^o(\rn)\,,\quad n\in[1,N]\,,
\label{cdm-2}
\end{equation}
whence
\begin{equation}
\+f^o(\rn) \approxeq \lec\*I-i\omega\*D_n\cdot\les \Copc-\Cop(\rn)\ris\ric^{-1}\cdot
\+f^o_{exc}(\rn)\,,\quad n\in[1,N]\,.
\label{cdm-3}
\end{equation}
Clearly, $\+f^o_{exc}(\rn)$ contains not only the incident optical field at $\rn$ but also
the scattered optical field due to all other lattice points in $\Vi$.

Substituting \r{cdm-3} in \r{cdm-1}, we obtain
\begin{equation}
\+f^o_h(\rn)=\sum_{m=1}^{N}\, \*Q^{CDM}_{nm}\cdot\+f^o_{exc}(\rm)\,,\qquad
n\in[1,N]\,,
\label{cdm-4}
\end{equation}
where
\begin{equation}
\label{Qcdm}
\*Q^{CDM}_{nm} =\*I\delta_{nm}-i\omega\vm\gopc(\rn,\rm)\cdot \mbox{\boldmath$\=\chi$}_m\,(1-\delta_{nm})\,
\end{equation}
contains
\begin{equation}
 \mbox{\boldmath$\=\chi$}_m=
\les\Copc-\Cop(\rm)\ris\cdot
\lec\*I-i\omega\*D_m\cdot\les \Copc-\Cop(\rm)\ris\ric^{-1}\,.
\end{equation}
Equations \r{cdm-4} have to be solved numerically too~---~just like \r{mom-2}~---~in order
to determine $\+f^o_{exc}(\rn)$, $n\in[1,N]$. Thereafter, by virtue of
\r{mom-3} and \r{cdm-3}, the
the total optical field at any $\#r\notin \Vi$ may be determined
from \r{6diffeqsol3} as
\begin{eqnarray}
&&\+f^o(\#r) \approxeq \+f^o_h(\#r)
+\sum_{m=1}^{N}\,
\vm\, \gopc(\#r,\rm)\cdot\mbox{\boldmath$\=\chi$}_m\cdot
\+f^o_{exc}(\rm)\,,\qquad\#r\notin \Vi\,.
\label{cdm-5}
\end{eqnarray}
The reason for the name of this technique is that the
product $\vm \mbox{\boldmath$\=\chi$}_m$ in
\r{Qcdm} and \r{cdm-5}
may be considered as a 6$\times$6 polarizability  dyadic of the electrically small
region associated with $\rm$ (Weiglhofer, Lakhtakia, \& Michel, 1997; Lakhtakia
\& Weiglhofer, 2000).

\subsection{Homogenization of Magnetostatic Field}
The coupled dipole method naturally leads to the homogenization
problem of replacing a spatially varying $\Bdc(\#r)$ in some region $\tilde{V}$   by
a uniform magnetostatic field $\#B_{uni}$ for the analysis of optical
fields in $\tilde{V}$.

Following the
strong--property--fluctuation theory (SPFT)
(Tsang \& Kong, 1981; Michel \& Lakhtakia, 1995; Mackay, Lakhtakia, \& Weiglhofer 2000), let us begin by introducing the continuous analog of
the exciting optical field of \r{cdm-3} as
\begin{equation}
 \+f^o_{exc} (\#r) = \lec \*I  - i\omega\*D\cdot \les \Copuni - \Cop(\#r)
 \ris \ric \cdot \+f^o(\#r)\,,\qquad \#r\in \tilde{V}\,,
\end{equation}
where $\Copuni$ is defined in analogy with $\Copc$~---~as per \r{epsopcdef}, \r{muopcdef},
and \r{Copcdef}~---~and $\*D$ is to be calculated by using $\Copc=\Copuni$ and $\=U=\=I$ on
the right side of \r{DD_def}.
Thereby, the integral equation \r{6diffeqsol3} may be rewritten as
\begin{equation}
 \+f^o_{exc} (\#r) = \+f^o_h(\#r)   + i\omega\,{\sf PV}\int_{\tilde V}  d^3\#r'\,
\gop_{uni} (\#r,\#r')\cdot \mbox{\boldmath$\=\chi$} (\#r') \cdot
\+f^o_{exc}(\#r')\, , \label{SPFT_ie}
\end{equation}
where $\gop_{uni} (\#r,\#r')$ is defined in analogy with $\gopc
(\#r,\#r')$ in \r{goc_def}  with $\Copc=\Copuni$, and the
polarizability density dyadic is defined as
\begin{eqnarray}
 \mbox{\boldmath$\=\chi$} (\#r)& =&
\les \Copuni- \Cop(\#r)\ris \cdot\lec \,\*I -
 i\omega\*D\cdot \les \Copuni- \Cop(\#r)\ris \ric^{-1}\,
\end{eqnarray}
in analogy to $ \mbox{\boldmath$\=\chi$}_m$.

Within the SPFT framework,
 the volume integrals of both sides of \r{SPFT_ie}
are used to compute the constitutive parameters of a  uniform medium
which  approximate to those of the medium described by $\Cop (\#r)$.
The simplest estimate of these constitutive
 parameters is provided by $\Cop_{uni}$, which is found by imposing the condition
\begin{equation}
\int_{\tilde V}d^3\#r \; \; \mbox{\boldmath$\=\chi$}(\#r) = \*0\,.
\l{chi0}
\end{equation}

In order to deduce $\#B_{uni}$ from \r{chi0}, we introduce the
piecewise--uniform approximation
\begin{equation}
\Bdc (\#r) = \#B_j, \qquad \#r \in \tilde{V}_j,
\end{equation}
with $\tilde{V} = \bigcup^J_{j=1} \tilde{V}_j$. The corresponding
uniform 6$\times$6 constitutive dyadics are defined as
\begin{equation}
\Cop (\#r) = \Copj = \les \begin{array}{cc}
\epso\,\epsopj & \=0\\[5pt]
\=0 & \muo\,\muopj
\end{array}\ris,  \qquad \#r \in \tilde{V}_j,
\end{equation}
with 3$\times$3 dyadic components
\begin{equation}
\left.
\begin{array}{l}
\epsopj = \le 1-8\epso\co^2\xi\,\#B_j \cdot \#B_j \ri\=I
+28\epso\co^2\xi\,\#B_j \#B_j \vspace{6pt} \\
\muopj = \les \le 1-8\epso\co^2\xi\,\#B_j \cdot \#B_j \ri\=I
-16\epso\co^2\xi\,\#B_j \#B_j \ris^{-1}
\end{array}
\right\}.
\end{equation}
Then, the condition \r{chi0} implies that
\begin{equation}
\sum_{j=1}^{J} \, f_j \le \Cop_{uni} - \Copj \ri \cdot \les \,\*I -
 i\omega\*D \cdot \le \Cop_{uni} - \Copj\ri \ris^{-1} = \*0\,,
\end{equation}
where $f_j$ is the volumetric proportion of region $\tilde{V}_j$
relative to $\tilde{V}$. This nonlinear dyadic equation can be
straightforwardly solved for $\Cop_{uni}$~---~and thereby
$\#B_{uni}$~---~ by standard numerical procedures (Michel,
Lakhtakia, \& Weiglhofer, 1998).

Higher--order corrections to $\Cop_{uni}$ can also be derived using the
SPFT, depending upon the statistical details of the spatial
fluctuations in $\Bdc (\#r)$ (Mackay, Lakhtakia, \& Weiglhofer, 2000).

\section{Concluding Remarks}
To conclude, we have formulated an integral equation for the scattering
of light (and other high--frequency) electromagnetic radiation in vacuum
by either a strong magnetostatic field or a strong low--frequency
magnetic field. This integral equation, set in the language of
6--vectors and 6$\times$6 dyadics, exploits the frequency--domain
dyadic Green function for a linear,  homogeneous, uniaxial, dielectric--magnetic medium.
The derived integral equation can be solved numerically,
using
 the method of moments as well as the coupled dipole method.
Finally, we have shown that the strong--property--fluctuation theory
allows the homogenization of a spatially varying magnetostatic field
in the present context.

\vspace{10mm}

\noindent{\bf Acknowledgement:} AL thanks John Collins of the
Physics Department, Penn State, for a discussion. TGM is supported by a \emph{Royal
Society of Edinburgh/Scottish Executive Support Research
Fellowship}.

\vspace{10mm}

 \noindent{\bf References}

Adler, S.L. 1971. Photon splitting and photon dispersion in a strong magnetic field.
\emph{Ann. Phys. (NY)} 67:599--647.

Adler, S.L. 2007. Vacuum birefringence in a rotating magnetic field.
\emph{J. Phys. A: Math. Theor.} 40:F143--F152.

Carnahan, B., H.A. Luther, and J.O. Wilkes. 1969. {Applied
Numerical Methods}. New York: Wiley.

Gradshteyn, I.S., and I.M. Ryzhik. 1980.  \emph{Table of Integrals,
Series, and Products}. New York: Academic Press.

Jackson, J.D. 1998. \emph{Classical Electrodynamics, 3rd edition}. New York:
Wiley (Section 1.3).

Lakhtakia, A. 1990. Macroscopic theory of the coupled dipole approximation method.
\emph{Opt. Commun.}  79:1--5.

Lakhtakia, A. 1992. Strong and weak forms of the method of moments and the
coupled dipole method for scattering of time--harmonic electromagnetic
fields. \emph{Int. J. Modern Phys. C} 3:583--603; corrections: 1993,
4:721--722.

Lakhtakia, A., and W.S. Weiglhofer. 2000. Maxwell Garnett formalism for weakly
nonlinear, bianisotropic, dilute, particulate composite media.
\emph{Int. J. Electron.} 87:1401--1408.

Lakhtakia, A., V.K. Varadan, and V.V. Varadan. 1991. Plane waves
and canonical sources in a gyro\-electro\-magnetic uniaxial medium.
\emph{Int. J. Electron.} 71:853--861.

Mackay, T.G., and A. Lakhtakia. 2005. Anisotropic enhancement of
group velocity in a homogenized dielectric composite medium. \emph{J. Opt.
A: Pure Appl. Opt.}  7:669--674.

Mackay, T.G., and W. S. Weiglhofer. 2000. Homogenization of biaxial
composite materials: dissipative anisotropic properties. \emph {J. Opt. A:
Pure Appl. Opt.}   2:426--432.

Mackay, T.G., A.   Lakhtakia, and W.S.   Weiglhofer. 2001.
Strong--property--fluctuation theory for homogenization of
bianisotropic composites: formulation. \emph{Phys. Rev. E}
62:6052--6064; corrections:  2001,  63:049901.

Michel, B., and A.  Lakhtakia. 1995. Strong--property--fluctuation
theory for homogenizing chiral particulate composites. \emph{Phys. Rev. E}
51:5701--5707.

Michel, B., and W. S. Weiglhofer. 1997.
Pointwise singularity of dyadic Green function in a
general bianisotropic medium. \emph{Arch. Elektron. \"Ubertrag.} 51:219--223;
correction: 1998, 52:31.

Michel, B.,  A.   Lakhtakia,  and W.S.    Weiglhofer. 1998. Homogenization of
linear bianisotropic particulate composite media~---~Numerical
studies. \emph{Int. J.  Appl.  Electromag.  Mech.}  9:167--178;
corrections:    1999, 10:537--538.

Miller, E.K., L. Medgyesi--Mitschang, and E. H. Newman. 1991. \emph{Computational
Electromagnetics: Frequency--Domain Method of Moments}. New York: IEEE Press.

Purcell, E.M., and C.R. Pennypacker. 1973. Scattering and absorption of light by
nonspherical dielectric grains. \emph{Astrophys. J.}   186:705--714.

Sarkar, T.K. (ed). 1991. \emph{Application of Conjugate Gradient Method to
Electromagnetics and Signal Analysis}. New York: Elsevier.

Singh, O.N., and A. Lakhtakia (eds). 2000. \emph{Electromagnetic Fields
in Unconventional Materials and Structures}. New York: Wiley.

Strang, G. 1986. \emph{Introduction to Applied Mathematics}.
Wellesley, MA: Wellesley--Cambridge Press.

Tsang, L., and J.A.   Kong. 1981.
Scattering of electromagnetic waves from random media with strong
permittivity fluctuations. \emph{Radio Sci.} 16:303--320.

Wang, J.J.H. 1991. \emph{Generalized Moment Methods in Electromagnetics}.
New York: Wiley.

Weiglhofer, W.S. 1990. Dyadic Green's functions for general uniaxial media.
\emph{IEE Proc., Pt. H}  137:5--10.

Weiglhofer,  W.S. 1998. Electromagnetic depolarization dyadics and
elliptic integrals. \emph{J. Phys. A: Math. Gen.} 31:7191--7196.

Weiglhofer, W.S., and A. Lakhtakia (eds). 2003. \emph{Introduction to Complex Mediums for Optics and Electromagnetics}. Bellingham, WA, USA: SPIE Press.

Weiglhofer, W.S., A. Lakhtakia, and B. Michel. 1997. Maxwell Garnett and
Bruggeman formalism for a particulate composite with bianisotropic
host medium. \emph{Microw. Opt. Technol. Lett.} 15:263--266;
correction: 1999, 22:221.

\end{document}